\documentclass[a4paper]{article}

\usepackage{fullpage}

\usepackage[mode=buildnew]{standalone}
\usepackage[utf8]{inputenc}
\usepackage{graphicx}
\usepackage{wrapfig}
\usepackage{amsmath}
\usepackage{amssymb}
\usepackage{enumerate}
\usepackage[hyphens]{url}
\usepackage{hyperref}
\usepackage{cleveref}
\usepackage[noend, ruled, plain, noline]{algorithm2e}
\usepackage{amsthm}
\DontPrintSemicolon
\usepackage{algpseudocode}
\usepackage{verbatim}

\usepackage{authblk}
\usepackage{todonotes}
\usepackage{tabularx}
\usepackage{microtype}
\usepackage{float}
\usepackage{tikz}
\usetikzlibrary{decorations.pathreplacing, shapes, calc}

\newcommand{\Oh}{\mathcal{O}}

\newcommand{\per}{\mathsf{per}}
\newcommand{\PREF}{\mathsf{Pref}}
\newcommand{\Occ}{\mathsf{Occ}}
\newcommand{\maxgap}{\mathsf{maxgap}}
\newcommand{\B}{\mathcal{B}}
\newcommand{\C}{\mathcal{C}}
\newcommand{\R}{\mathcal{R}}

\newcommand{\AFI}{\textbf{AFI}\xspace}
\newcommand{\Folk}{\textbf{Folk}\xspace}
\newcommand{\Bres}{\textbf{Bres}\xspace}
\newcommand{\MS}{\textbf{MS}\xspace}

\newcommand{\IMP}{\textbf{IMP}\xspace}
\newcommand{\KKRRWs}{\textbf{KKRRW-s}\xspace}
\newcommand{\KKRRW}{\textbf{KKRRW}\xspace}

\newtheorem{theorem}{Theorem}[section]

\newtheorem{lemma}[theorem]{Lemma}
\newtheorem{observation}[theorem]{Observation}

\theoremstyle{remark}
\newtheorem{definition}[theorem]{Definition}

\newtheorem{example}[theorem]{Example}

\begin{document}

\title{Experimental Evaluation of Algorithms for Computing Quasiperiods}
\author{
Patryk Czajka}
\author{Jakub~Radoszewski}

\affil{
Institute of Informatics, University of Warsaw, Warsaw, Poland\\
\texttt{patryk.czajka@student.uw.edu.pl, jrad@mimuw.edu.pl}}
\date{\vspace{-.5cm}}

\maketitle              
\begin{abstract}
Quasiperiodicity is a generalization of periodicity that was introduced in the early 1990s.
Since then, dozens of algorithms for computing various types of quasiperiodicity were proposed.
Our work is a step towards answering the question: ``Which algorithm for computing quasiperiods to choose in practice?''.
The central notions of quasiperiodicity are covers and seeds.
We implement algorithms for computing covers and seeds in the original and in new simplified versions
and compare their efficiency on various types of data.
We also discuss other known types of quasiperiodicity, distinguish partial covers as currently the most
promising for large real-world data, and check their effectiveness using real-world data.

\end{abstract}

\section{Introduction}
Quasiperiodicity was introduced by Apostolico and Ehrenfeucht~\cite{DBLP:journals/tcs/ApostolicoE93}
as an extension of standard periodicity.
The most basic type of a quasiperiod is a \emph{cover}.
We say that a string $C$ is a cover of a string $S$ if every position of $S$ lies within an occurrence of $C$.
A generalization of notion of cover is the notion of \emph{seed}.
A seed is a cover of a superstring of $S$.
In other words, we allow the seed to cover positions of $S$ with overhanging occurrences.
For example, the shortest cover of string $aabaaabaabaa$ is $aabaa$ and the shortest seeds of this string are $aaba$ and $abaa$.

An $\Oh(n)$-time algorithm that computes the shortest cover of a string of length $n$ was given
by Apostolico et al.\ \cite{DBLP:journals/ipl/ApostolicoFI91}.
An $\Oh(n)$-time on-line algorithm for the same problem was proposed by Breslauer~\cite{DBLP:journals/ipl/Breslauer92}.
Moore and Smyth \cite{DBLP:journals/ipl/MooreS94,DBLP:journals/ipl/MooreS95} proposed a linear-time algorithm computing all the covers of a string; its simpler implementation can be inferred from a recent work of Crochemore et al.~\cite{DBLP:journals/tcs/CrochemoreIKRRW17}.
Finally, Li and Smyth \cite{DBLP:journals/algorithmica/LiS02} developed a linear-time algorithm for computing the length of the longest cover of every prefix of a string.
The output of their algorithm can be used to compute all the covers of any prefix of the string.

Seeds were introduced and first studied by Iliopoulos et al.~\cite{DBLP:journals/algorithmica/IliopoulosMP96}
who proposed an $\Oh(n \log n)$-time algorithm computing a representation of all the seeds in a string of length $n$.
A different algorithm with the same time complexity computing the shortest seed of a string was proposed by Christou et al.~\cite{DBLP:journals/tcs/ChristouCIKPRRSW13};
they also showed how to fill in a gap in the algorithm of Iliopoulos et al.~\cite{DBLP:journals/algorithmica/IliopoulosMP96}.
Finally, a linear-time algorithm for computing all the seeds was presented by Kociumaka et al.~\cite{DBLP:conf/soda/KociumakaKRRW12};
it was simplified in a very recent technical report~\cite{DBLP:journals/corr/abs-1107-2422}.

Due to applications in molecular biology, both covers and seeds have been extended into the case of
covering a string with multiple quasiperiods \cite{IliopoulosSmythKCovers}.
This way, the notions of
$k$-covers \cite{DBLP:journals/jalc/ColeIMSY05,DBLP:journals/isci/IliopoulosMS11},
$\lambda$-covers \cite{DBLP:journals/isci/GuoZI07}, and
$\lambda$-seeds \cite{DBLP:conf/aaim/GuoZI06} were introduced.
In applications such as molecular biology and computer-assisted music
analysis, finding exact repetitions is not always sufficient; the same problem
occurs for quasiperiodic repetitions.
This led to the introduction of the notions of approximate covers and seeds
\cite{DBLP:journals/algorithmica/AmirLLLP19,DBLP:conf/cpm/AmirLLP17,DBLP:conf/cpm/AmirLP18,DBLP:conf/itat/GuthMB08,DBLP:journals/jalc/ChristodoulakisIPS05,SimParkKimLee},
enhanced covers, partial covers and seeds \cite{DBLP:journals/tcs/FlouriIKPPST13,GUTH2019,DBLP:journals/algorithmica/KociumakaPRRW15,DBLP:journals/tcs/KociumakaPRRW16},
and approximate $\lambda$-covers~\cite{DBLP:conf/spire/GuoZI06}.

Other results in this area include computation of covers and seeds in the
parallel~\cite{DBLP:conf/soda/Ben-AmramBIP94,DBLP:journals/ipl/Breslauer94,Korea}
and streaming models~\cite{StreamingCovers}.
Another line of research is finding maximal quasiperiodic substrings of a string, for which
an $\Oh(n\log^2{n})$-time algorithm~\cite{DBLP:journals/tcs/ApostolicoE93}
and $\Oh(n\log{n})$-time algorithms~\cite{DBLP:conf/cpm/BrodalP00,DBLP:journals/jalc/IliopoulosM99}
were proposed.
Intermediate variants of quasiperiodicity between covers and seeds called left and right seeds were also considered
\cite{DBLP:journals/jda/ChristouCGIP12,DBLP:journals/tcs/ChristouCIKPRRSW13,DBLP:journals/tcs/FlouriIKPPST13},
as well as recovering a string from the set of covers of its prefixes~\cite{DBLP:conf/cpm/CrochemoreIPT10}.
Combinatorial properties of covers were presented
in~\cite{DBLP:journals/ipl/AmirIR17,DBLP:journals/arscom/ChristouCI16,DBLP:journals/tcs/IliopoulosM99}.
Covers were also considered in indeterminate \cite{DBLP:journals/dam/AlatabbiRS16,DBLP:conf/stringology/AntoniouCIJL08,DBLP:journals/tcs/CrochemoreIKRRW17,DBLP:journals/njc/IliopoulosMMPST03} and weighted
strings~\cite{DBLP:journals/corr/BartonKLPR17,DBLP:conf/cpm/BartonKPR16,DBLP:journals/fuin/IliopoulosMPPTT06}.
A survey by Apostolico and Breslauer~\cite{DBLP:conf/birthday/ApostolicoB97}
describes in detail the algorithms~\cite{DBLP:journals/tcs/ApostolicoE93,DBLP:journals/ipl/Breslauer92,DBLP:journals/ipl/Breslauer94,Korea}, whereas
a survey by Iliopoulos and Mouchard~\cite{DBLP:journals/tcs/IliopoulosM99} describes the
algorithms~\cite{DBLP:journals/tcs/ApostolicoE93,DBLP:journals/ipl/ApostolicoFI91,DBLP:journals/algorithmica/IliopoulosMP96}.
Early results on quasiperiodicity were also surveyed by Smyth~\cite{DBLP:journals/tcs/Smyth00}.

\paragraph{\bf Our contributions.}
\begin{itemize}
\item We provide implementations of known efficient algorithms for computing quasiperiods (covers, seeds) and approximate quasiperiods (partial covers)
and compare their efficiency using extensive computer experiments.
\item We show how to use a data structure called Joinable Segment Trees to obtain an alternative $\Oh(n \log n)$-time version of the
linear-time algorithm for computing seeds that turns out to perform superior on practical data.
The same data structure is used for efficient computation of partial covers.
\item We perform a detailed review of the literature on quasiperiodicity and approximate quasiperiodicity, identify
approaches which are currently most promising for computations on large practical data, and additionally point out a flaw
in one of the known algorithms (computation of $\lambda$-covers).
\end{itemize}

\paragraph{\bf Structure of the paper.}
In Section~\ref{sec:covers} we discuss in detail known linear-time algorithms for computing covers
\cite{DBLP:journals/ipl/ApostolicoFI91,DBLP:journals/ipl/Breslauer92,DBLP:journals/ipl/MooreS94,DBLP:journals/ipl/MooreS95,DBLP:journals/tcs/CrochemoreIKRRW17} and a folklore $\Oh(n \log n)$-time algorithm and
compare their efficiency on the most basic problem of computing the shortest cover.
For this, we use random data as well as semi-random data that is guaranteed to have non-trivial covers.
In Section~\ref{sec:seeds} we make the same type of comparison of algorithms for computing seeds: the $\Oh(n \log n)$-time algorithm
\cite{DBLP:journals/algorithmica/IliopoulosMP96}, the simplified version \cite{DBLP:journals/corr/abs-1107-2422}
of the $\Oh(n)$-time algorithm~\cite{DBLP:conf/soda/KociumakaKRRW12} and a simpler $\Oh(n \log n)$-time version
of~\cite{DBLP:journals/corr/abs-1107-2422}.
Here the algorithms are much more involved, so their descriptions are necessarily rather sketchy.
Finally, in Section~\ref{sec:general} we discuss approximate variants of string quasiperiodicity, distinguish partial covers
as currently the most promising on large real-world data, and test the $\Oh(n \log n)$-time algorithm for partial
covers~\cite{DBLP:journals/algorithmica/KociumakaPRRW15} both on artificial data and data from
Pizza\&Chili corpus\footnote{\texttt{http://pizzachili.dcc.uchile.cl}}.
For the $\Oh(n \log n)$-time implementations
of the algorithms from~\cite{DBLP:journals/corr/abs-1107-2422,DBLP:journals/algorithmica/KociumakaPRRW15},
we use a compact segment tree data structure which is especially well suited in practice that we describe in Appendix~\ref{sec:JST}.

The C++ programs that were used for comparisons are available at \url{https://github.com/heurezjusz/Quasiperiods}.
Our experiments were conducted on Intel® Core™ i7-7700HQ processor with 16GB RAM.

\section{Preliminaries}

The letters of a string $T$ are numbered $1$ through $|T|$.
By $T[i]$ we denote the $i$-th letter of $T$ and by $T[i,j]$ we denote $T[i] \dots T[j]$ which we call
a substring of $T$.
For strings $T$ and $X$, we denote $\Occ_T(X) = \{i\,:\,T[i,i+|X|-1]=X\}$.

We say that an integer $p$ is a \emph{period} of a string~$T$ if $T[i]=T[i+p]$
for all $i=1,\ldots,|T|-p$.
By $\per(T)$ we denote the smallest period of $T$.
The string $T$ is called \emph{periodic} if $2 \, \per(T) \le |T|$.
We say that a string~$B$ is a \emph{border} of $T$ if $B$ is both a prefix and a suffix of $T$.
The border array $\B$ of $T$ stores, as $\B[i]$, the length of the longest proper border of $T[1,i]$.
It can be computed in linear time using the Knuth--Morris--Pratt (KMP) algorithm \cite{DBLP:journals/siamcomp/KnuthMP77}.

For a set of integers $A=\{a_1,\ldots,a_k\}$, $a_1 < a_2 < \dots < a_k$, by $\maxgap(A)$ we denote the maximum
distance between consecutive elements of $A$: $\maxgap(A) = \max\{a_i-a_{i-1}\,:\,i=2,\ldots,k\}$.
If $|A| \le 1$, we assume that $\maxgap(A) = \infty$.

\begin{definition}
A string $C$ is a \emph{cover} of a string $T$ if $$\maxgap(\Occ_T(C)\cup\{-|C|+1,|T|+1\})=|C|.$$ A string $S$ is a \emph{seed} of $T$ if $|S| \le |T|$ and $S$ is a cover of some string containing $T$ as a substring.
\end{definition}

\section{Computing covers}\label{sec:covers}

In this section we compare algorithms for computing covers.
They are all relatively simple comparing to the algorithms for computing seeds, so we can afford to explain them
with more details.
We start by two algorithms which only compute the shortest cover of a string, a folklore one
and the algorithm of Apostolico et al.\ \cite{DBLP:journals/ipl/ApostolicoFI91}, proceed to the algorithm
of Breslauer~\cite{DBLP:journals/ipl/Breslauer92} which computes the shortest cover of every prefix of the string and
finish by the algorithm by Moore and Smyth \cite{DBLP:journals/ipl/MooreS94,DBLP:journals/ipl/MooreS95}
which computes all covers of a string.
We decided not to consider the algorithm of Li and Smyth~\cite{DBLP:journals/algorithmica/LiS02}
that computes the longest cover array of a string since it is much more involved.
Yet another linear-time algorithm for computing the shortest cover can be inferred as a special case of the linear-time algorithm for computing so-called enhanced
covers~\cite{DBLP:journals/tcs/FlouriIKPPST13}; see Section~\ref{sec:general}.
We also did not consider this algorithm in the following comparison.

The first two algorithms are based on a relation between covers and borders.

\begin{observation}\label{obs:cover_basic}
Let $T$ be a string.
\begin{enumerate}[(a)]
  \item\label{a} A cover of $T$ is a border of $T$.
  \item\label{b} If $T$ has a cover $C$ and a border $B$ such that $|C| \le |B|$, then $C$ is also a cover of $B$.
  \item\label{c} The shortest cover of~$T$ is not periodic.
\end{enumerate}
\end{observation}

A string is called \emph{superprimitive} if it is equal to its shortest cover, and \emph{quasiperiodic} otherwise.

\subsection{Computing shortest cover}
We describe two algorithms that compute only the shortest cover of a string, a folklore $\Oh(n \log n)$-time algorithm
and the $\Oh(n)$-time algorithm of Apostolico et al.\ \cite{DBLP:journals/ipl/ApostolicoFI91}.
Let $b_1 < \dots < b_k$ be the sequence of lengths of all borders of $T$.
By Observation~\ref{obs:cover_basic}\eqref{a}, these are all candidates for a cover of $T$.
Both algorithms are based on a folklore fact that the sequence $b_1,\ldots,b_k$ can be partitioned
into at most $\log n$ subsequences, each of which is an arithmetic sequence.
Moreover, if any of these arithmetic sequences has at least 3 elements and difference $p$, then
all the borders in this subsequence excluding possibly the first one are periodic with period $p$.

The algorithms use the fact that testing if a string $S$ is a cover of a string $T$ can be done in linear time
by computing the set $\Occ_T(S)$ using, say, the KMP algorithm and then applying the operation $\maxgap$.
The border array can also be used to compute all the borders of a string in linear time.

In the first algorithm we notice that if $b_i \le 2 b_{i-1}$, then the border of length $b_i$ is periodic.
Hence, it cannot be the shortest cover by Observation~\ref{obs:cover_basic}\eqref{c}.
The number of the remaining indices $i$, for which $b_i > 2 b_{i-1}$, does not exceed $\log_2 n$.
Hence, we can check each of them as a candidate for the cover length.
Let us call this algorithm \Folk.

The algorithm of Apostolico et al.\ \cite{DBLP:journals/ipl/ApostolicoFI91}, further denoted as \AFI, is recursive.
It starts by computing the longest border $B$ of $T$.
If $|B|\le \frac23 n$, it recursively finds the shortest cover of this border and
checks whether it is a cover of $T$. If it is, the shortest cover of the border is returned as an answer, and otherwise the result is the whole $T$.
Correctness of this step follows by Observation~\ref{obs:cover_basic}\eqref{b}.
Otherwise, $B$ is periodic and instead of $B$ it takes the longest shorter border $B'$ of $T$ that is not periodic
with the same period.
We have $|B'| = p + n \bmod{p}$, where $p=n-|B|$, so $|B'| \le \frac23 n$.
The recurrence on the time complexity, $T(n) = T(\frac23 n)+\Oh(n)$, yields $T(n) = \Oh(n)$.

\subsection{Computing shortest cover on-line}
The algorithm of Breslauer~\cite{DBLP:journals/ipl/Breslauer92}, further denoted as \Bres,
is arguably the most popular algorithm for computing covers.
It computes the shortest cover of every prefix of the string in an on-line manner.

The algorithm makes use of three arrays $\B$, $\C$, $\R$, where $\B$ is the border array, $\C[i]$ is the length of the shortest cover of $T[1,i]$, and $\R[i]$ is not defined if $T[1,i]$ is not superprimitive and otherwise stores the length of the longest prefix of $T$, up to the latest arrived letter $T[k]$, such that $T[1,i]$ is its cover. In the end, $\C[n]$ contains the length of the shortest cover of $T$. See the pseudocode below.

\begin{algorithm}[h!]
\For{$k:=1$ \KwSty{to} $n$}{
	\If{$\B[k] > 0$ \KwSty{and} $\R[\C[\B[k]]] \ge k - \C[\B[k]]$}{
	  $\C[k] := \C[\B[k]]$\;
    $\R[\C[k]] := k$
  }\Else{
		$\C[k] := \R[k] := k$
  }
}
\end{algorithm}

  In the if-statement of the algorithm we check if the shortest cover of the border $T[1,\B[k]]$ covers $T[1,k]$.
  If so, then it is the shortest cover of $T[1,k]$ by Observation~\ref{obs:cover_basic}\eqref{b}.
  Otherwise, $\B[k] = 0$ or $\C[\B[k]]$ does not cover $T[1,k]$, so $T[1,k]$ is superprimitive.

\subsection{Computing all covers}\label{sec:cov3}
The algorithm of Moore and Smyth \cite{DBLP:journals/ipl/MooreS94,DBLP:journals/ipl/MooreS95}
computes all the covers of a string $T$.
We describe its slightly simplified version that can be inferred from the work of
Crochemore et al.~\cite{DBLP:journals/tcs/CrochemoreIKRRW17}.

The algorithm creates a list $L$ of all positions $\{1,\ldots, n\}$ with an additional element $n+1$.
It performs $n$ steps.
In step $i$ it removes from $L$ all positions $j$ which are not starting positions of occurrences of $T[1,i]$.
Hence, after step $i$ the list $L$ represents $\Occ_T(T[1,i])$.
The algorithm stores the $\maxgap$ of $L$.
If after the $i$th phase the $\maxgap$ of remaining elements is smaller than or equal to $i$, then $T[1,i]$ is a cover of $T$.
The $\maxgap$ of $L$ can only increase, since the elements 1 and $n+1$ are never removed.
When an element is removed from $L$, the $\maxgap$ is maximized with the difference of its next and previous elements in $L$.
Elements can be removed efficiently with the aid of an additional array that stores the position of each element from $\{1,\ldots,n+1\}$ in $L$.

We only need to specify how to identify the elements of the list that are not occurrences of $T[1,i]$.
The original algorithm~\cite{DBLP:journals/ipl/MooreS95} used a data structure called a \emph{border tree}.
A border tree is a rooted tree that contains a node for each position $i\in \{0,\ldots,n\}$ in $T$.
The parent of $i$ is the node $B[i]$.
We chose a different, more direct solution~\cite{DBLP:journals/tcs/CrochemoreIKRRW17} that uses the
\emph{prefix array} $\PREF$ which holds, as $\PREF[j]$, the length $\ell$ of the longest prefix of $T$
such that $T[1,\ell]=T[j,j+\ell-1]$.
The prefix array can be computed in $\Oh(n)$ time by a simple left-to-right scan~\cite{Jewels}.
Then position $j$ should be removed from the list $L$ at step number $\PREF[j]+1$.

We denote the resulting algorithm as \MS.

\subsection{Experimental evaluation}
As it was mentioned in the beginning of this section, the algorithms serve different purposes.
We compare their efficiency on the fundamental task of computing the shortest cover.
Hence, this comparison is necessarily limited as it does not convey their different capabilities.

For a string $S$, an \emph{equality relation} states that a given pair of substrings, $S[i,i+\ell]$ and $S[j,j+\ell]$, match.
Such relation is specified by a triple $(i,j,\ell)$.
In \cite{DBLP:conf/wads/GawrychowskiKRR15} a linear-time algorithm is presented that, given a collection of equality relations,
constructs a string of a specified length over the largest possible alphabet that satisfies these relations.
We used a simpler, $\Oh(n \log n)$-time version of this algorithm that was also described
in \cite{DBLP:conf/wads/GawrychowskiKRR15} to generate quasiperiodic strings with a given set of occurrences of a cover.

\paragraph{\bf General conclusions.}
Graphs that present the running time of the algorithms can be found in Appendix~\ref{sec:covers1}.
The best running times on the tests was achieved by the algorithms \Folk and \AFI,
both achieving similar times. Performance of on-line algorithm \Bres
depended on string properties.
The \MS algorithm, capable of computing all covers of a string, was the slowest.

\paragraph{\bf Two versions of \MS algorithm.}
Because the algorithm computes a lot of auxiliary values, we compared its straightforward
implementation with a second version that uses static arrays instead of dynamic \texttt{std::vector}. It is visible that
static arrays take noticeable time for initialization, but it pays off in tests with longer strings.

\paragraph{\bf Periodicity of the string.}
For this and the following paragraphs, see the graphs in Appendix~\ref{sec:covers2}.
The algorithm \Bres slows down for small periods. One can notice that the slowdown seems to
be linear in border length. The cause might be that in case of finding a cover of some prefix $T[1,i]$
shorter than $i$, which happens quite often for short periods,
the algorithm performs more write operations. The smaller the period is,
the sooner the algorithm finds the cover which repeats with every period.
This might create the effect visible on the plot.

The running times of algorithms \Folk and \AFI clearly depend on the period.
An interesting split point
is when the period reaches $\frac{n}{2}$. The cause is probably that the \AFI algorithm
performs linear-time checks on prefixes of the input string, while the $\Oh(n\log n)$-time algorithm checks cover candidates
on the whole string. The difference shows up when the input string has a short border.

Also there is a noticeable slowdown of the \MS algorithm for short periods (i.e.\ 2, 3).
The reason is that for such strings the $\PREF$ array achieves the largest values, and computation
time of $\PREF$ depends on its maximum value.

\paragraph{\bf Alphabet size.}
The running time does not seem to depend on the alphabet size.
For small alphabets, all approaches have slightly worse execution times,
but the reason is that strings consisting of less different letters are more likely to
have some kind of regularity, which, as the experiments have shown, increases the execution times.

\paragraph{\bf Shortest cover length.}
There is no clear dependency on the cover length. Fluctuations of performance suggest
that the running time probably depends on other properties of the string (like periodicity)
caused by the existence of a cover.

\section{Computing seeds}\label{sec:seeds}
We consider three algorithms for computing seeds in a string of length $n$: an $\Oh(n \log n)$-time algorithm by
Iliopoulos et al.~\cite{DBLP:journals/algorithmica/IliopoulosMP96}, an $\Oh(n)$-time
algorithm of Kociumaka et al.~\cite{DBLP:journals/corr/abs-1107-2422} and its simpler $\Oh(n \log n)$-time version.

Each of the three algorithms computes a linear representation of all seeds of the input string $T$.
The representation is a set of packages. A single package is specified with three integers $i$, $j_1$, $j_2$
and represents strings $T[i,j_1]$, $T[i,j_1+1]$, \ldots, $T[i,j_2]$. For example, for $T=aabcab$, the package $(2,4,5)$
represents two strings: $abc$ and $abca$. Kociumaka et al.~\cite{DBLP:journals/corr/abs-1107-2422}
prove that all seeds of $T$ can be represented as a linear number of such packages.

Consider an edge of the suffix tree of $T$, connecting two nodes $u$ (parent) and $v$ (child), denoted as $e_{uv}$.
Let us denote by $|w|$ the depth of node $w$, which also is the length of the substring represented by this node.
Note that all substrings represented by the edge $e_{uv}$ (excluding $u$ and including $v$) can be represented by a single package
$(i_{suf}, i_{suf}+|u|+1, i_{suf}+|v|)$, where $i_{suf}$ is the starting position of the suffix of $T$ represented by any leaf
in the subtree of the node $v$.

The algorithm of Iliopoulos et al.~\cite{DBLP:journals/algorithmica/IliopoulosMP96}
also uses packages on the reversed string $T$.
An efficient way to convert between reversed and usual package representations is not known.

The main difference in the algorithms lies in how the $\maxgap$ values are computed for sets that represent
substrings of the string $T$ with the same occurrences.

\subsection{$\Oh(n \log n)$-time algorithm of Iliopoulos et al.~\cite{DBLP:journals/algorithmica/IliopoulosMP96}}

Let us call this algorithm \IMP.
The algorithm uses Crochemore's partitioning;
see \cite{DBLP:journals/ipl/Crochemore81} (see also~\cite[Section 9.1]{DBLP:books/daglib/0020103}).
The partitioning performs $n$ steps;
on the $i$-th step it maintains $\Occ_T(S)$ as a sorted list for all $i$-length substrings $S$ of $T$.
The total time complexity of partitioning is $\Oh(n\log n)$.

Iliopoulos et al.\ show that every substring of $T$ of length greater than or equal to $\per(T)$ is a seed of $T$.
Their algorithm computes $\per(T)$ and reports $\Oh(\per(T))$ packages with seeds longer than or equal to it.
Then it performs $\per(T)$ steps of Crochemore's partitioning to find all shorter seeds.

\newcommand{\pos}{\mathit{pos}}

Iliopoulos et al.\ use the notion of a \textit{candidate set}. A candidate set consists
of substrings $W_{i_0}$, $W_{i_1}$, \ldots, $W_{i_l}$ of $T$ such that $|W_{i_j}|=i_j=i_0+j$ and
$\Occ_T(W_{i_j})=\Occ_T(W_{i_0})=\{\pos_1, \pos_2,\ldots,\pos_k\}$ for all $j=0,\ldots,l$.
Note that a candidate set can be represented as a single package $(\pos_1, \pos_1+i_0-1, \pos_1+i_l-1)$.
The authors show how to determine the set of seeds from a candidate set in $\Oh(1)$ time,
using $\pos_1$, $\pos_k$, $i_0$, $i_l$, $\maxgap({\Occ_T(W_{i_0})})$ and a number of auxiliary arrays that are
precomputed in $\Oh(n)$ time using the KMP algorithm. Due to the auxiliary arrays the algorithm needs to be run twice --
once on reversed $T$, what implies using reversed packages on the output.

The candidate sets can be found by Crochemore's partitioning.
Let us assume that a single $\Occ_T(S)$ list was created on step $i_0$ and was partitioned into smaller
lists in step $i_l$. It represents a candidate set of substrings $W_{i_0}$, $W_{i_1}$, \ldots, $W_{i_l}$.
Note that in Crochemore's partitioning, after elements from some list have been removed,
the list does not have to be empty. In this case the remaining elements represent $\Occ_T(W)$ for some string $W$.
This is the second way the $\Occ_T$ lists are created in Crochemore's partitioning.

The candidate set can be computed from $\Occ_T(W_{i_l})$ list just before it will be partitioned
into smaller lists. The positions $\pos_1$ and $\pos_k$ can be found trivially, because the list is sorted. We
can extend the list by the moment of its creation to find $i_0$.
The problem occurs when we want to extend the list by its $\maxgap$, which will be updated
in $\Oh(1)$ time when some element is removed. The problem is easy when one can assume that the smallest
and the largest element are never removed from the list, as it was the case in the algorithm
of Moore and Smyth \cite{DBLP:journals/ipl/MooreS94,DBLP:journals/ipl/MooreS95} that was described in Section~\ref{sec:cov3}.
Unfortunately, it does not have to be the case during the partitioning.

For this issue, Christou et al.~\cite{DBLP:journals/tcs/ChristouCIKPRRSW13} proved that, in case of seeds computation,
one can skip removing extreme elements from the list without losing correctness. Thus $\maxgap$ can
be easily maintained, simply by not updating the $\maxgap$ value when an extreme element is removed.

\subsection{$\Oh(n \log n)$-time version of the algorithm of Kociumaka et al.~\cite{DBLP:journals/corr/abs-1107-2422}}

The paper operates on the notions of \textit{a left candidate}, \textit{a right candidate} and \textit{a quasiseed}.
A string $S$ is a \textit{left candidate} if for some $i$ it is a suffix and a seed of $T[1,i]$.
A string $S$ is a \textit{right candidate} if for some $i$ it is a prefix and a seed of $T[i,n]$.
Finally, a string $S$ is a \textit{quasiseed} if $\maxgap(\Occ_T(S)) \le |S|$.
Kociumaka et al.\ prove that a string $S$ is a seed if $T$ if and only if it fulfills all three above definitions.

The paper shows that sets of left and right candidates can be found in $\Oh(n)$ time
and returned in a package representation. It also presents a method of finding an intersection of
sets of substrings represented as packages in linear time.
The algorithm does not use reversed packages. Also no seeds in the result are duplicated.

The original method for computing quasiseeds in $\Oh(n)$ time is described in Section~\ref{below}.
Here we describe a simpler method working in $\Oh(n\log n)$ time that we used in the first implementation, called \KKRRWs.

Let us denote as $e_{uv}$ an edge of the suffix tree of $T$ connecting the nodes $u$ (parent) and $v$ (child).
All substrings of $T$ represented by this edge have the same $\Occ_T$ -- they occur at the beginning
of all suffixes of $T$ represented by leaves in the subtree of $v$. 
The package containing all quasiseeds of $T$ represented by $e_{uv}$ is
$(i_{suf}, i_{suf}+\max(|u|+1,\maxgap(\Occ_T(v))), i_{suf}+|v|)$, where $i_{suf}$
represents a starting position of any suffix of $T$ in the subtree of $v$.
Obviously if $\maxgap(\Occ_T(v)) > |v|$, the edge $e_{uv}$ does not contain any quasiseeds.

To find all quasiseeds of $T$, one needs to compute $\maxgap(\Occ_T(v))$
for every node $v$. In our implementation we used Joinable Segment Trees.
Implementation details are described in Appendix~\ref{sec:JST}.
A similar data structure was introduced in~\cite{DBLP:conf/swat/Karczmarz16}.

\subsection{$\Oh(n)$-time algorithm of Kociumaka et al.~\cite{DBLP:journals/corr/abs-1107-2422}}\label{below}

We denote this algorithm as \KKRRW.
The difference between the algorithms \KKRRWs and \KKRRW lies in $\maxgap$ computation.

Kociumaka et al.\ prove a \textit{Gap Lemma} which states that:
\begin{lemma}
If a string $T$ has at least $\frac{2}{3}n-k+1$ substrings of length $k$, then $T$ has no seed
of length $\ell$ such that $2k-2\le \ell\le \frac{n}{6}$.
\end{lemma}

Note that ``at least $\frac{2}{3}n-k+1$ substrings of length $k$'' can be described also as
``at least $\frac{2}{3}n$ substrings of length $k$ including all prefixes of $T$ of length smaller than $k$'';
the latter is further denoted as $\beta_k(T)$. One can easily conclude
that $\beta_i(T) \le \beta_{i+1}(T)$ for all $i$.

The algorithm computes the maximum $x$ such that $\beta_{4x-3}(T)<\frac{2}{3}n$
(it can be done in $\Oh(n)$ time using the suffix tree of $T$).
Then it computes all seeds of $T$ divided into three groups according to the Gap Lemma:
\emph{long seeds}, of length $\ell > \frac{n}{6}$, in $\Oh(n)$ time;
\emph{medium seeds}, of length $x < \ell < 2k-2 = 2(4x-3)-2<8x$, in $\Oh(n)$ time;
and short seeds, of length $\ell< x$, recursively on a string of length $\frac{2}{3}n$.

\paragraph{\bf Long seeds.} These are the seeds of length $\ell > \frac{n}{6}$.
In the Gap Lemma the upper limit on $\ell$ does not depend on $k$.
We can use this fact to count all long seeds in $\Oh(n)$ time.
Left and right candidates are computed in the same way as in the $\Oh(n\log n)$-time version of the algorithm.
In the computation of quasiseeds, we compute $\max(\maxgap(\Occ_T(v)),\frac{n}{6}+1)$
for every node $v$ of the suffix tree of $T$ and then deduce the packages describing quasiseeds
for every edge $e_{uv}$ basing on the values in $v$.
The intersection of these three package representations
can be computed in $\Oh(n)$ time.

Thanks to the fact that we want to compute only quasiseeds of length greater than $\frac{n}{6}$,
the algorithm computing $\maxgap(\Occ_T(v))$ values for every node $v$ is much simpler.
We can split the list of positions $1,\ldots,n$ into 6 blocks of equal lengths
and for each block remember the first and the last occurrence of $v$ in $T$.
Values $\maxgap(\Occ_T(v))$ that are greater than $\frac{n}{6}$ can be computed correctly
from the 12 values remembered in the 6 blocks. We can create and update such a block structure in $\Oh(1)$ time,
thus the total work necessary to compute long seeds is $\Oh(n)$.

\paragraph{\bf Medium seeds.} These are the seeds of length $x < \ell < 2k-2 = 2(4x-3)-2<8x$.
In this case, a fact that a string $S$ is a seed of $T$ if and only if
it is a seed of every substring of $T$ of length $2|S|-1$ is used.
Kociumaka et al.\ show that it is sufficient to compute seeds of
substrings of $T$ of length $16x$ with step $8x$, that is $T[1,16x]$, $T[8x+1,24x]$, $T[16x+1,32x]$ and so on.
The total length of the selected substrings is $2n$.
Note that we are looking for seeds of length at least $x$, so we can use similar
method as in the computation of long seeds, but with 16 blocks.
Intersection of solutions found of all such substrings can be computed in $\Oh(n)$ time,
thus the total work necessary to compute the medium seeds is $\Oh(n)$.

\paragraph{\bf Short seeds.} Finally, these are the seeds of length $\ell < x$.
Using the fact that was mentioned in the previous paragraph, we will find recursively
seeds of all substrings of $T$ of length $2x-1$ and compute an intersection of the results.

The algorithm first marks all positions covered by first occurrences
of all substrings of $T$ of length $2x-1$ (the starting positions of these occurrences can be read from the suffix tree of $T$).
Then it recursively computes the seeds of strings created by maximal intervals of the marked positions.
Note that every marked position $p$ either is the middle letter of the first occurrence of a string
$T[p-2x+1,p+2x-1]$ of length $4x-3$ or is contained in the prefix $T[1,2x]$. Because $\beta_{4x-3} < \frac{2}{3}n$, the total number
of marked positions is at most $\frac{2}{3}n$.

\bigskip
Total work of the algorithm is limited by $c \cdot n +c\cdot\frac{2}{3}n+c\cdot\frac{2^2}{3^2}n+\ldots=c\cdot 3n=\Oh(n)$,
where $c$ is a constant.

\subsection{Experimental Evaluation}
We used Ukkonen's construction~\cite{DBLP:journals/algorithmica/Ukkonen95} of the suffix tree.
The children of a node were indexed in a hash map by the letters of the alphabet.

The experimental evaluation (see Section~\ref{sec:seeds1}) shows that the $\Oh(n \log n)$-time \KKRRWs
algorithm worked faster on experimental data than its original $\Oh(n)$ version \KKRRW. Performance of the \IMP algorithm
depended on the type of the string.
Low performance of the \KKRRW algorithm is probably a result of a large constant.
\footnote{As an example, computing so-called medium seeds in the \KKRRW algorithm requires computing 16 values for every node of a suffix tree.
  Overall, we have estimated that the linear-time algorithm has a chance of overtaking its $\Oh(n\log n)$
  version on strings of length around $2^{50}$. Unfortunately, available equipment has not allowed us to check
  this hypothesis.
}

On random data, the \KKRRWs algorithm had the best performance. The other two
algorithms achieved similar times. On the other hand, on periodic strings
the \IMP algorithm becomes slightly better.
The reason behind this is probably that the performance of the \IMP algorithm strongly
depends on the period and the shortest period of a random string is not likely to be smaller than $n$.

We checked dependencies of algorithm's performance on multiple properties of a string.
The most interesting results are listed below.

\paragraph{\bf String period.} The plot clearly shows a linear dependency of the \IMP algorithm
on period length. For periods longer than $\frac{n}{2}$ it starts to behave like for random strings.
The dependency of \KKRRW algorithm is really interesting -- it shows for which
periods it starts to make recursive calls. We can see that the plot suddenly drops for a period
about $\frac{2}{3}n$, which is the border for the first recursive call.
The rule for subsequent recursive calls is more complicated, thus it is harder to predict
the positions of smaller peaks.

\paragraph{\bf Alphabet size.} (See Section~\ref{sec:seeds2}.) We expected that the running times of all algorithms would not depend on the alphabet size.
The outcome shows that the \IMP and \KKRRW algorithms performs slower for a smaller alphabet.
For \IMP the reason is the Crochemore's partitioning mechanism. For a smaller alphabet, when a list
is split into shorter lists, it produces on average a smaller number of lists that are longer, which results in more splits overall.
\KKRRW slows down for a similar reason as for algorithms computing the shortest cover
-- strings consisting of a smaller number of letters are more likely to have some regularity, which slows down this solution.

Experiments do not show any dependency of shortest seed length on algorithms' performance.
The huge fluctuations on the plot are a result of other string properties, which differed between tests.

\section{Approximate and Generalized Quasiperiodicity}\label{sec:general}

\subsection{Comparison of Known Approaches}
While quasiperiodicity was introduced as a generalization of periodicity~\cite{DBLP:journals/tcs/ApostolicoE93},
the notions of covers and seeds can still be too restrictive in discovering repeating patterns in strings.
To address this issue, approximate and generalized notions of quasiperiodicity were introduced.

If one allows just a portion of positions of the string to be covered, the notion of a \emph{partial cover} is obtained.
More precisely, we seek for all substrings of the string that cover at least a given number $\alpha$ of positions.
E.g., $aba$ is a partial cover of $abababbaba$ which covers $8$ positions of the string.
Partial covers can be computed fast, in $\Oh(n \log n)$ time~\cite{DBLP:journals/algorithmica/KociumakaPRRW15}.
If a string has a partial cover that covers $\alpha$ positions, then a maximum set of non-overlapping occurrences of this partial
cover covers at least $\alpha/2$ positions.
Hence, finding a substring with a maximum number of non-overlapping occurrences~\cite{DBLP:conf/icalp/BrodalLOP02}
gives a good approximation of a partial cover.
Another notion, of an \emph{enhanced cover}, requires the partial cover to be a border of the string.
Enhanced covers can be computed in $\Oh(n)$ time~\cite{DBLP:journals/tcs/FlouriIKPPST13}, but at the same time
they are more restrictive.
A generalization of partial covers are \emph{partial seeds}, in which letters covered by overhanging occurrences
are also counted.
They can also be computed in $\Oh(n \log n)$ time~\cite{DBLP:journals/tcs/KociumakaPRRW16}.
The gain in comparison with partial covers takes place only near the ends of a string.

One can also consider the case that a given string $S$ does not have an exact cover, but it is
at a small distance from a string $T$ that has an exact cover.
This way, we obtain the notion of an \emph{approximate cover}, as the shortest cover of a string $T$ that has the minimum
Hamming distance from a given string $S$.
E.g., $aba$ is an approximate cover of $abababbaba$ because it is an exact cover of $ababaababa$ that is at Hamming distance 1 from the string.
Unfortunately, the problem of computing an approximate cover of a string is NP-hard~\cite{DBLP:conf/cpm/AmirLLP17}.
Several relaxations of this problem were considered.
If one knows the set of positions where the approximate cover occurs in the string $T$, then
it can be computed in linear time~\cite{DBLP:conf/cpm/AmirLLP17,DBLP:conf/cpm/AmirLP18}.
Unfortunately, this does not help if we know just the string $S$.
Assuming that the approximate cover has at least one exact occurrence in the given string $S$, it can be computed
in polynomial time.
Unfortunately, the fastest known algorithm works in $\Omega(n^3)$ time~\cite{DBLP:journals/algorithmica/AmirLLLP19},
which is unrealistic for big real-world data.
An NP-hard version of approximate covers was also studied in \cite{SimParkKimLee}.

Yet another definition allows to cover the string with a family of strings that are all similar to one string
that is considered an \emph{approximate cover}.
For the similarity measure, usually the Hamming distance is considered and a parameter $k$ that is the maximum allowed distance is specified.
In \cite{DBLP:conf/itat/GuthMB08} an $\Oh(n^3k)$-time solution to the approximate cover problem with an additional assumption
that the approximate cover is a substring of the given string is shown.
Furthermore, \cite{GUTH2019} extended this definition in a natural way to approximate enhanced covers and proposed an equally fast algorithm for solving this problem
(if, in addition, the approximate enhanced cover needs to be a border of the given string, it is shown that the problem can be solved in $\Oh(n^2)$ time).
In the variant that allows overhanging occurrences, we obtain a notion of \emph{approximate seeds} that was considered
in~\cite{DBLP:journals/jalc/ChristodoulakisIPS05}.
If the approximate seed has an exact occurrence in the string, this work proposed an $\Oh(n^3)$-time algorithm for variants of the Hamming distance
and an $\Oh(n^4)$-time algorithm for different variants of the edit distance.
However, in the general case it is also shown that this problem is NP-complete.

A different line of research is obtained if one is looking for, instead of one cover, a family of $\lambda$ strings,
all of the same length $k$, whose occurrences cover the given string $S$.
Depending on which of the parameters is specified, we obtain the problem of $\lambda$-covers or $k$-covers; the other
parameter is to be minimized.
E.g., $abababbaba$ has a cover $(ab,ba)$ with $\lambda=k=2$.
The $k$-covers problem is NP-complete~\cite{DBLP:journals/jalc/ColeIMSY05,DBLP:journals/isci/IliopoulosMS11}.
In \cite{DBLP:journals/isci/GuoZI07} the authors claim that the $\lambda$-covers problem
can be solved in $\Oh(n^2)$ time for a constant $\lambda$ and a constant-sized alphabet.
The analogously defined $\lambda$-seeds problem was considered in \cite{DBLP:conf/aaim/GuoZI06}, where an $\Oh(n^2)$-time
algorithm is proposed, and a problem of approximate $\lambda$-covers was considered in~\cite{DBLP:conf/spire/GuoZI06},
where an $\Oh(n^4)$-time algorithm is proposed; in both cases $\lambda$ and the size of the alphabet were assumed to be constant.
In \cite{DBLP:journals/isci/GuoZI07} it is stated that the algorithm actually computes
all the sets of $\lambda$ strings of equal length that cover the given string, under
additional constraints that $1 < k < n/\lambda$ and that no proper subset of the $\lambda$ strings covers the string $S$.
However, the example below shows that the number of $\lambda$-covers under such definition can still be $\Omega(n^{\lambda})$,
so they cannot be computed in $\Oh(n^2)$ time.

\begin{example}
  Let the alphabet be $\{a,b_1,\ldots,b_{\lambda-1}\}$, for a constant $\lambda$, and consider the string
  $$S=(a^m b_1)^2 (a^m b_2)^2 \ldots (a^m b_{\lambda-1})^2 a^m.$$
  Then for every non-negative integers $i_1,j_1,\ldots,i_{\lambda-1},j_{\lambda-1}$ such that
  $$0 < i_1+j_1 = \ldots = i_{\lambda-1}+j_{\lambda-1} < m,$$
  $(a^{i_1} b_1 a^{j_1},\ldots,a^{i_{\lambda-1}} b_{\lambda-1} a^{j_{\lambda-1}},a^{i_1+j_1+1})$ forms a $\lambda$-cover of $S$ consisting of strings of length $i_1+j_1+1$.
  The number of such $\lambda$-covers is $\Omega(m^\lambda) = \Omega(|S|^{\lambda})$.
  Indeed, for every $\frac{m}{2} \le i_1+j_1 < m$, there are at least $\frac{m}{2}$ options for each of the parameters
  $i_1,\ldots,i_{\lambda-1}$.
\end{example}

In conclusion, we have decided to implement and test the algorithm for partial covers, as a good example of a known algorithm for computing approximate quasiperiodicity
with fast running time.

\subsection{Partial Covers}

In this problem we are to find all shortest substrings of a string
$T$ whose occurrences cover at least $\alpha$ positions of $T$.


The paper~\cite{DBLP:journals/algorithmica/KociumakaPRRW15} provides the definition of a \textit{Cover Suffix Tree} (CST) of $T$,
which is a suffix tree of $T$ where implicit nodes representing
primitively rooted squares of $T$ are converted to explicit ones.
String $U$ is \emph{primitive}, if for any string $V$ and integer $k$, $V^k=U$ implies $V=U$.
\textit{A primitively rooted} square is a string $U^2$, where $U$ is \textit{primitive}.
By known combinatorial results on squares in strings, this way only a linear number of nodes will be added.

The paper shows how to construct a CST of $T$ in $\Oh(n\log n)$ time with its nodes annotated by
two values: $cv(v)$ and $\Delta(v)$; $cv(v)$ is the number of positions of $T$ covered by occurrences
of the string represented by node $v$ and $\Delta(v)$ is the number of maximal fragments of $T$
that are fully covered by occurrences of $v$. More formally,
$\Delta(v)=1+|\{i\in \Occ_T(v): j-i > |v|$, $j$ is a successor of $i$ in $\Occ_T(v)\}|$.

Consider an edge $e_{uv}$ of CST, connecting two explicit nodes $u$ (parent) and $v$ (child).
All implicit nodes on $e_{uv}$ have the same value $\Delta$, which equals $\Delta(v)$.
This means that for an implicit node $w$, $\Delta(w)=\Delta(v)$ and $cv(w)=cv(v)-(|v|-|w|)\Delta(v)$.
Because $cv$ values of all nodes on edge $e_{uv}$ form an arithmetic sequence, it is easy to determine
whether on $e_{uv}$ there is a node $w$ such that $cv(w) \ge \alpha$ and find all such nodes with minimal depth.

This problem can be extended to \textit{All Partial Covers} problem in which for each $\alpha \in \{1,\ldots,n\}$
we are to find any shortest substring of $T$ whose occurrences cover at least $\alpha$ positions of $T$.
Consider an edge $e_{uv}$ containing $k$ implicit nodes. We can represent it
as a segment on $\mathbb{N}^2$ plane connecting points $(|v|, cv(v))$ and $(|v|-k, cv(v)-k\Delta(v))$.
The upper envelope of such segments can be found in $\Oh(m\log m)$ time, where $m$ is the number of segments.
Then point $(x,y)$ of such envelope represents that $y$ positions can be covered by a substring of length $x$.
One can find examples of such substrings by labeling segments by edges id.
This solution of \textit{All Partial Covers} problem works in $\Oh(n\log n)$ time.

One of subproblems that appear in computing partial covers is an extended Disjoint Set Union problem,
in which we require that a Union operation on sets $A$ and $B$ returns a \textit{change list}.
A change list consists of pairs $(x, next[x])$, where $x \in (A \cup B)$ and $next[x]$ is a successor of $x$ in this set, such that
$x$ is included in a change list if and only if the value $next[x]$ changed as a result
of the union. For example, the change list of the union of sets $\{1,2,3,5\}$ and $\{4,8\}$
is $[(3,4),(4,5),(5,8)]$.
The paper~\cite{DBLP:journals/algorithmica/KociumakaPRRW15} solves this problem in $\Oh(n\log n)$ time
using mergeable AVL trees.
An equally efficient solution of this problem with JSTs that we used in our implementation is described in Appendix~\ref{JST:FU}.

\subsection{Experimental Evaluation}

The experiments on Pizza\&Chili corpus have shown that in most cases the string which covers the maximum number of positions
of the text is a letter which occurs most of the times. The original algorithm computes the shortest partial cover for a given $\alpha$, so such a letter overwrote
all other partial covers. We slightly changed the algorithm to report all partial covers (not only the shortest one)
covering at least $\alpha$ positions, but the results were similar to ones achieved for random strings.
The only shortest partial covers longer than one letter were detected in XML source: "\texttt{journal}" covering 7.5\% of the text and
"\texttt{></article><articlemdate="200}" covering 7.6\% of the text.
However, both of them are border-free and this corresponds to a non-overlapping partial cover.

\section{Conclusions}
The main purpose of this work was a practical and, to some extent, theoretical comparison of algorithms for computing various
types of quasiperiods in strings.
From our study we conclude that the area of approximate quasiperiodicity seems to require further study in order to develop approaches that are applicable in practice.
We hope that this work will also be beneficial as a reference material on the state of the art in this area.

\paragraph{Acknowledgements.}
The authors thank Tomasz Kociumaka and Juliusz Straszyński for helpful discussions.

\bibliographystyle{plainurl}
\bibliography{main}

\appendix

\section{Supplementary graphs for Section~\ref{sec:covers}}

\subsection{Performance dependency on length of different kinds of strings}\label{sec:covers1}

\begin{center}
	\includegraphics[scale=0.27]{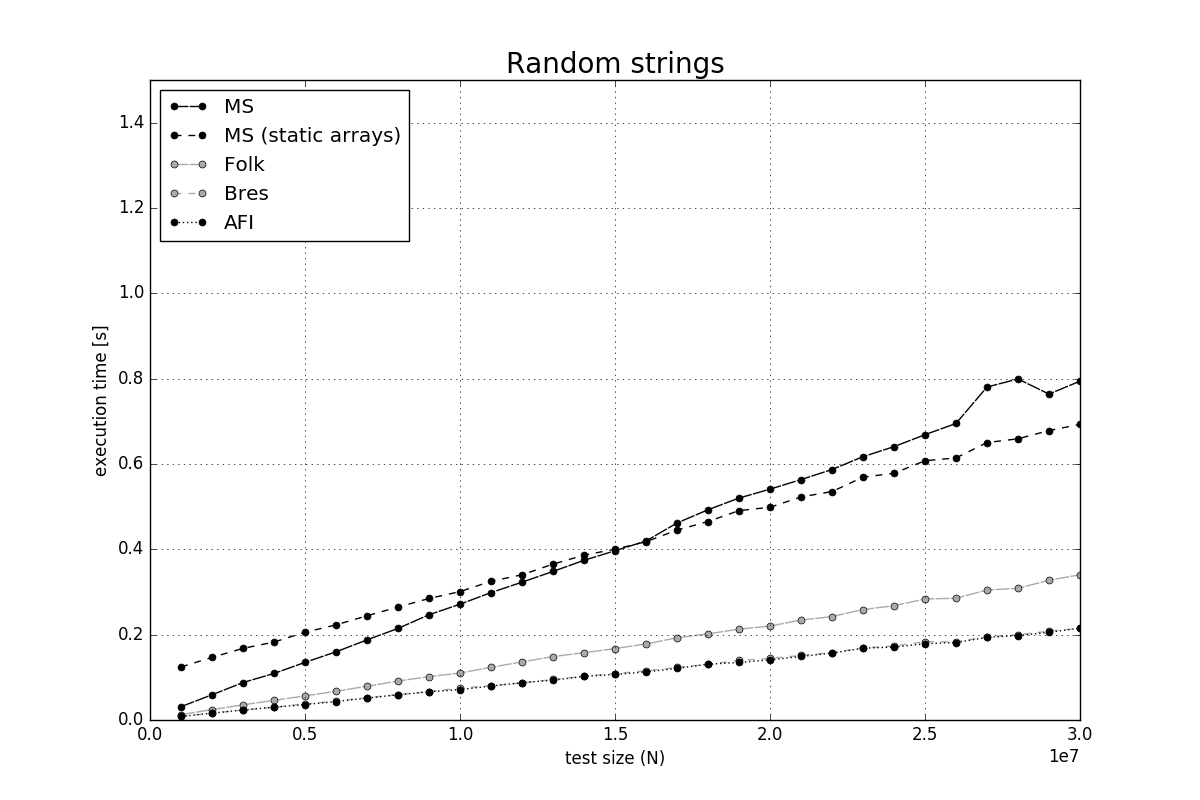}
\end{center}
\vspace*{-0.3cm}

\begin{center}
  \includegraphics[scale=0.27]{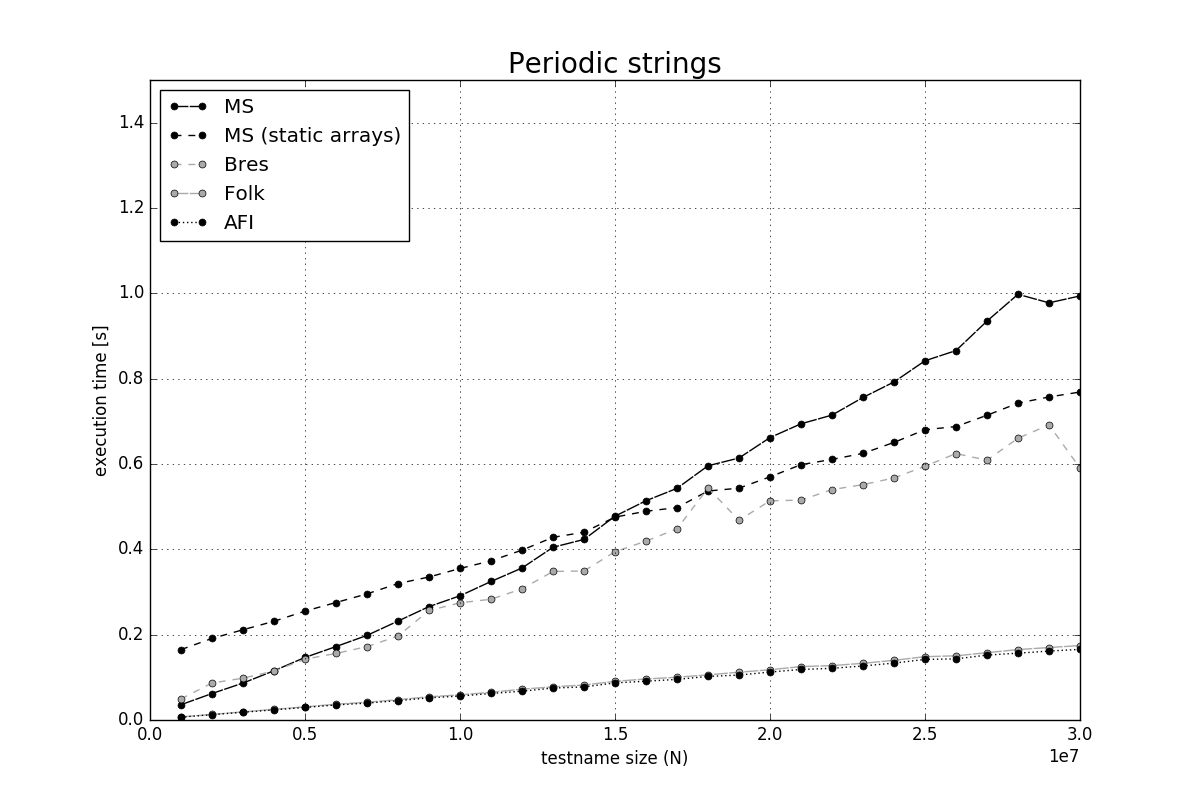}
\end{center}
\vspace*{-0.3cm}

\begin{center}
  \includegraphics[scale=0.27]{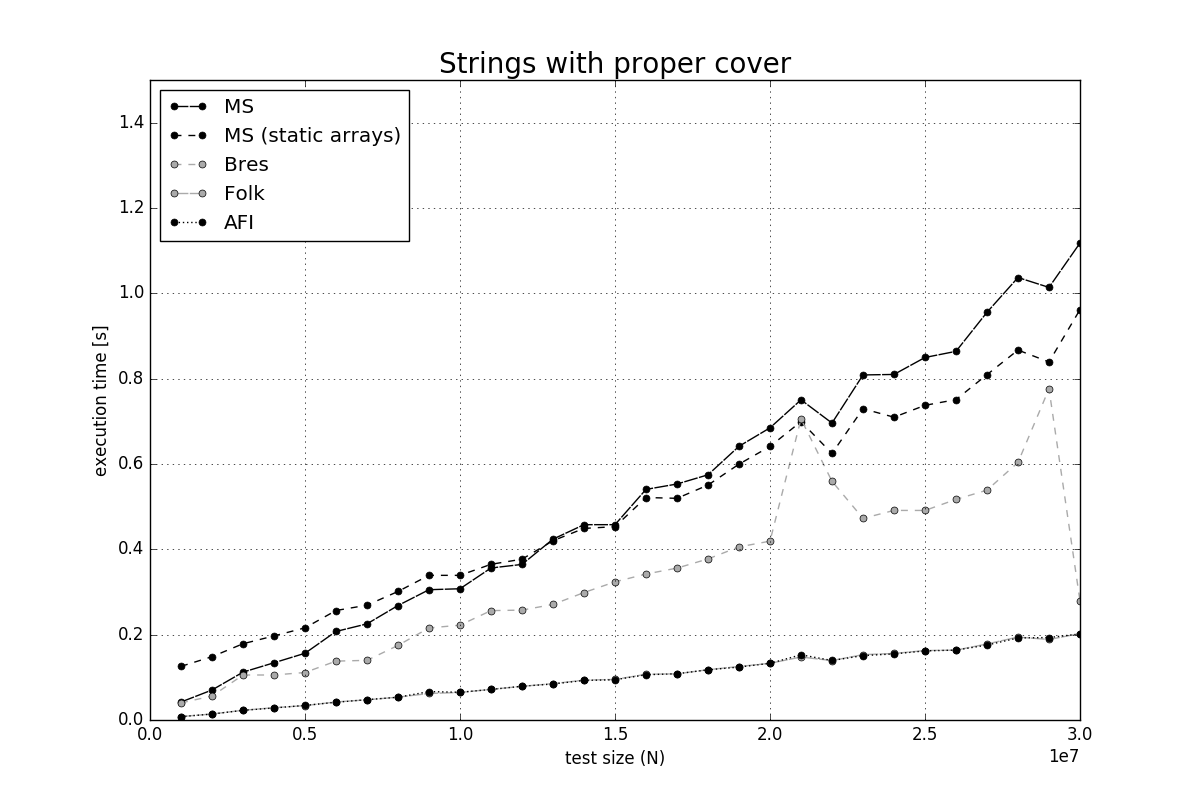}
\end{center}
\vspace*{-0.3cm}

\subsection{Performance dependency on different properties of the string}\label{sec:covers2}

\begin{center}
	\includegraphics[scale=0.3]{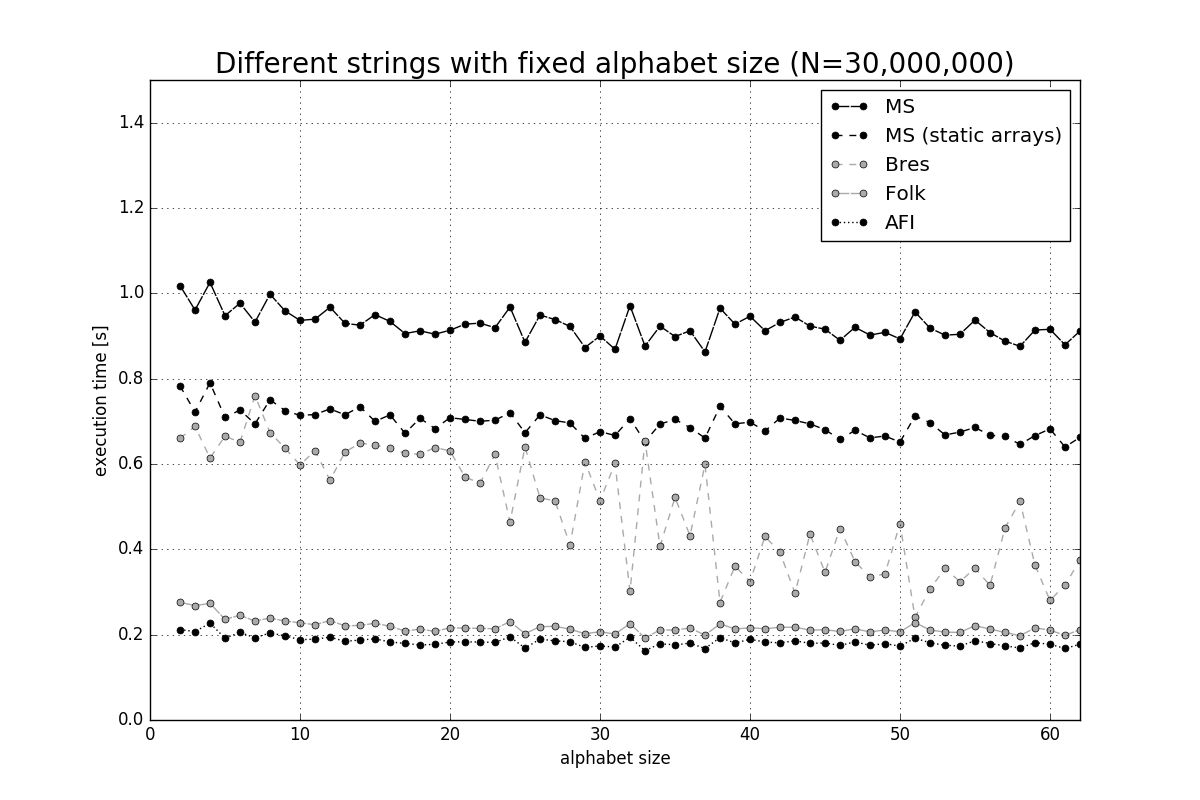}
\end{center}
\vspace*{-0.3cm}

\begin{center}
  \includegraphics[scale=0.3]{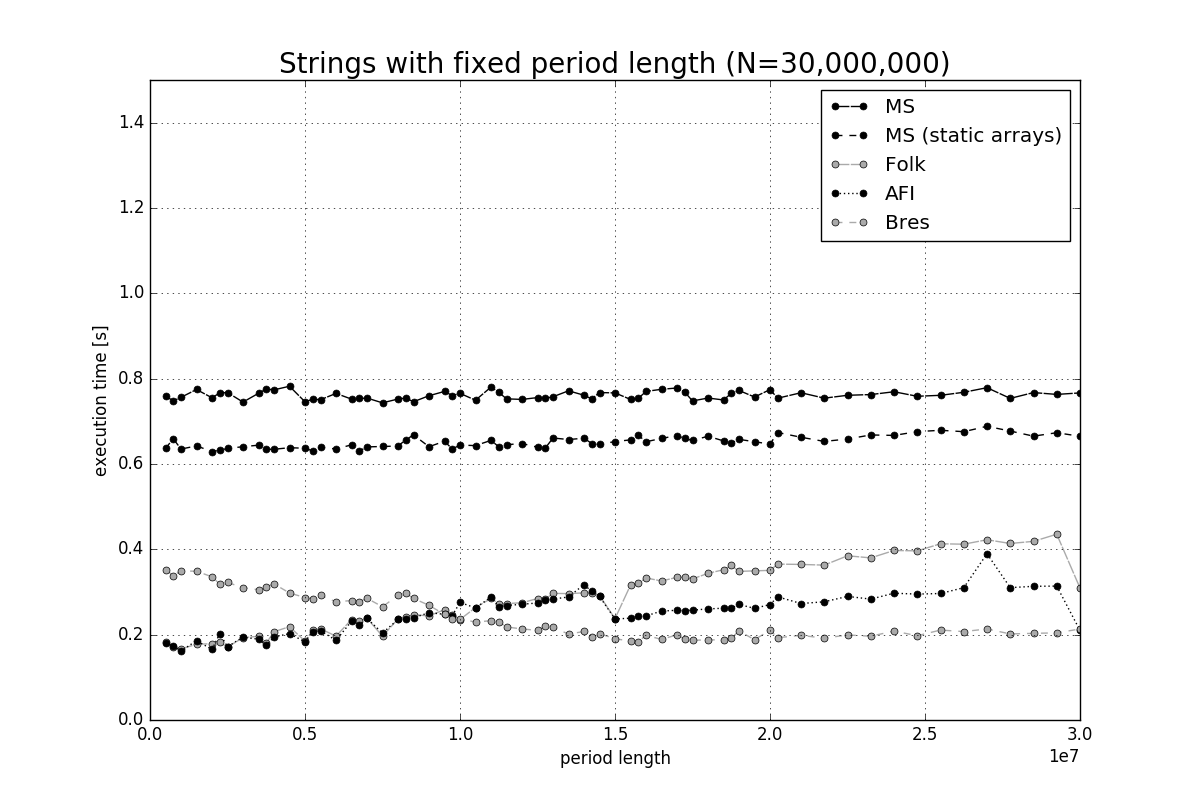}
\end{center}
\vspace*{-0.3cm}

\begin{center}
	\includegraphics[scale=0.3]{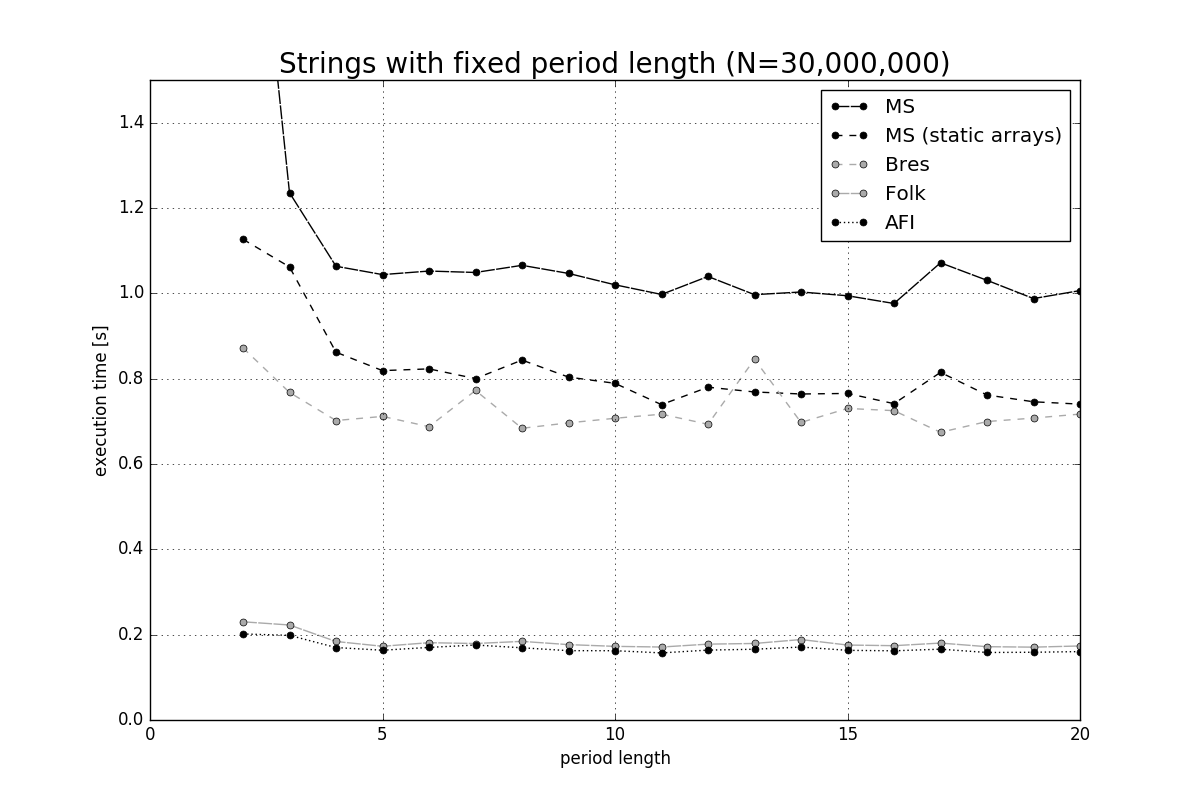}
\end{center}
\vspace*{-0.3cm}

\begin{center}
  \includegraphics[scale=0.3]{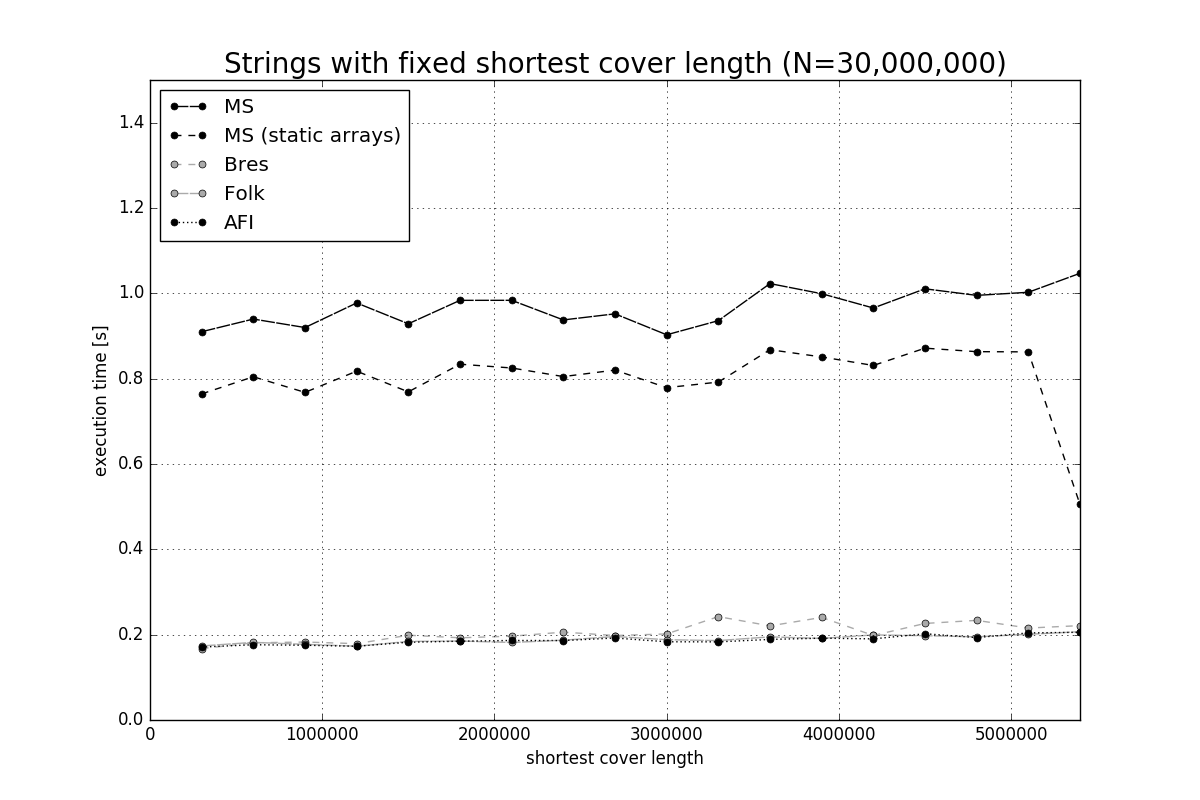}
\end{center}

\section{Supplementary graphs for Section~\ref{sec:seeds}}

\subsection{Performance dependency on length of different kinds of strings}\label{sec:seeds1}

\begin{center}
	\includegraphics[scale=0.27]{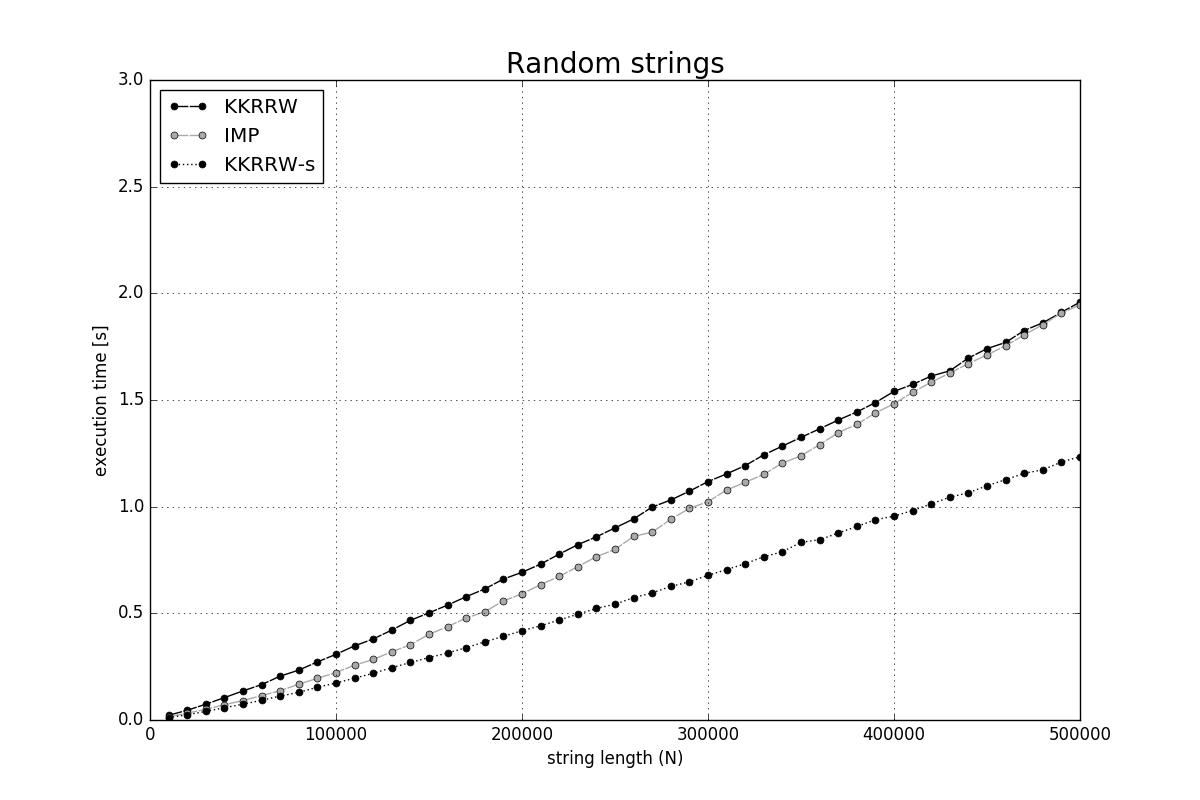}
\end{center}
\vspace*{-0.3cm}

\begin{center}
	\includegraphics[scale=0.27]{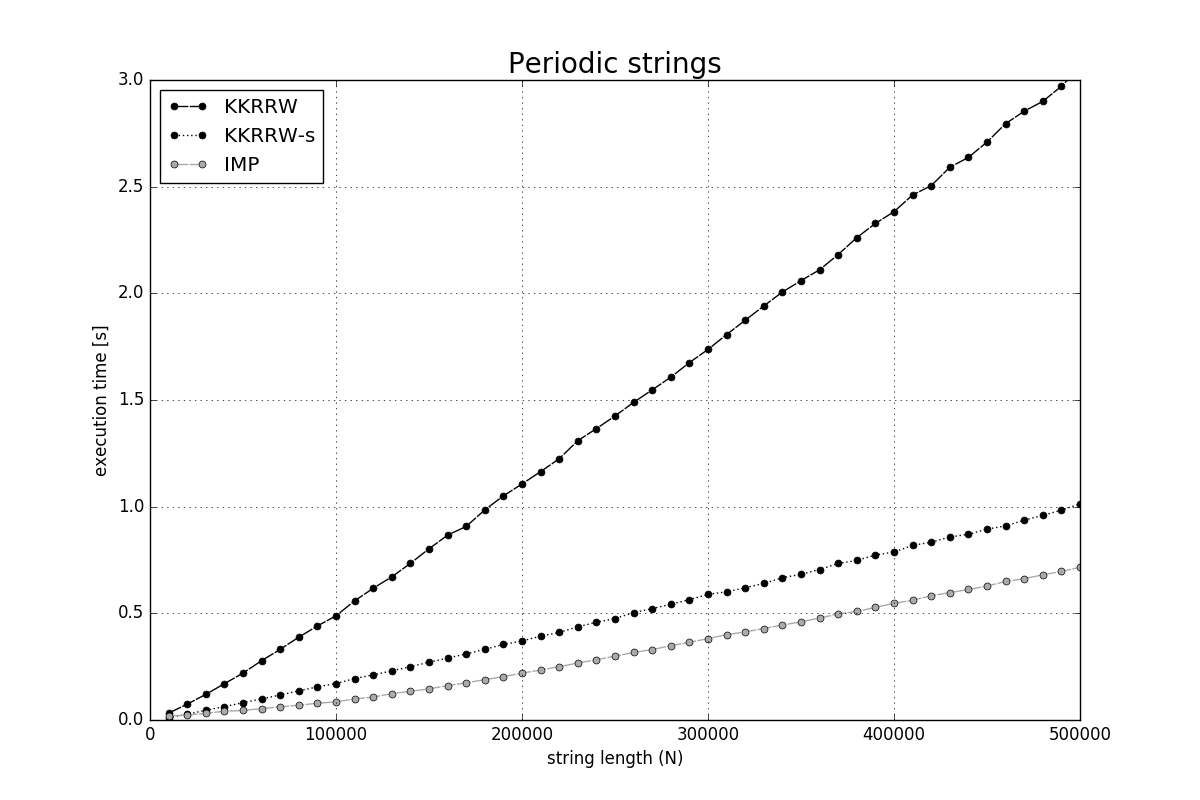}
\end{center}
\vspace*{-0.3cm}

\begin{center}
	\includegraphics[scale=0.27]{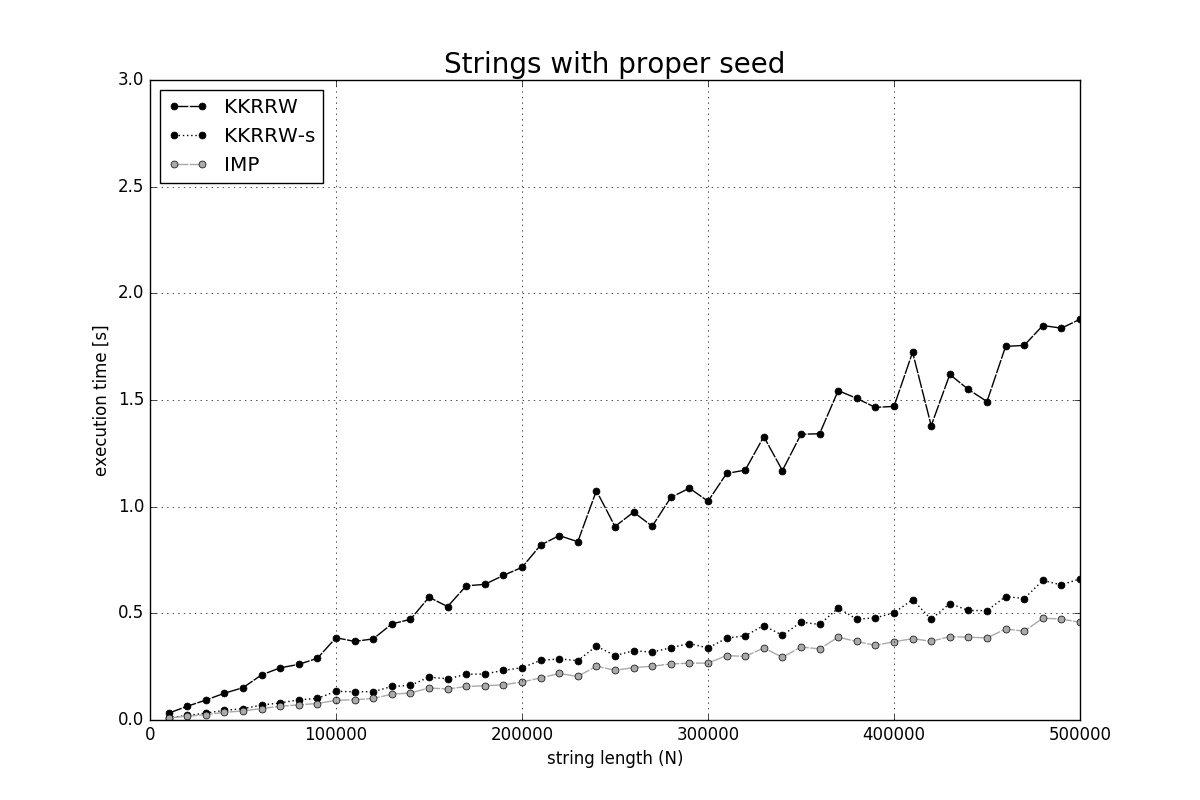}
\end{center}
\vspace*{-0.3cm}

\subsection{Performance dependency on different properties of the string}\label{sec:seeds2}

\begin{center}
	\includegraphics[scale=0.27]{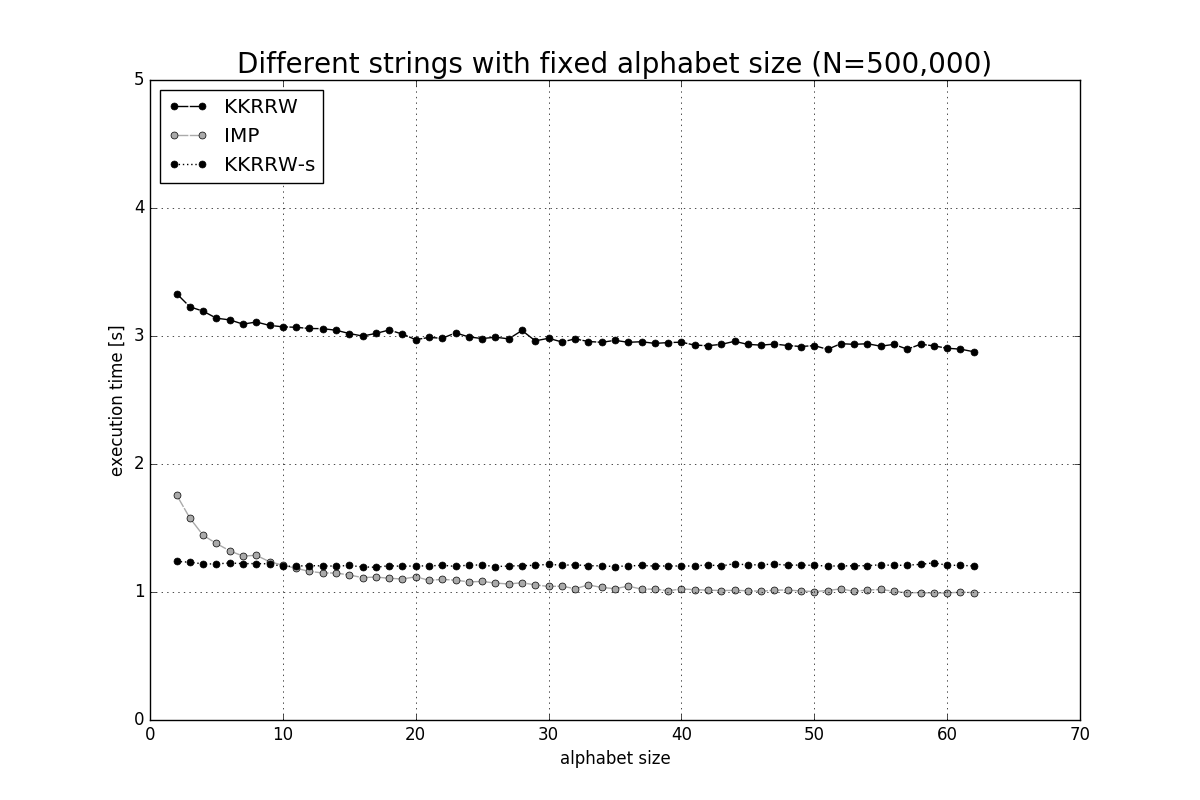}
\end{center}
\vspace*{-0.3cm}

\begin{center}
	\includegraphics[scale=0.27]{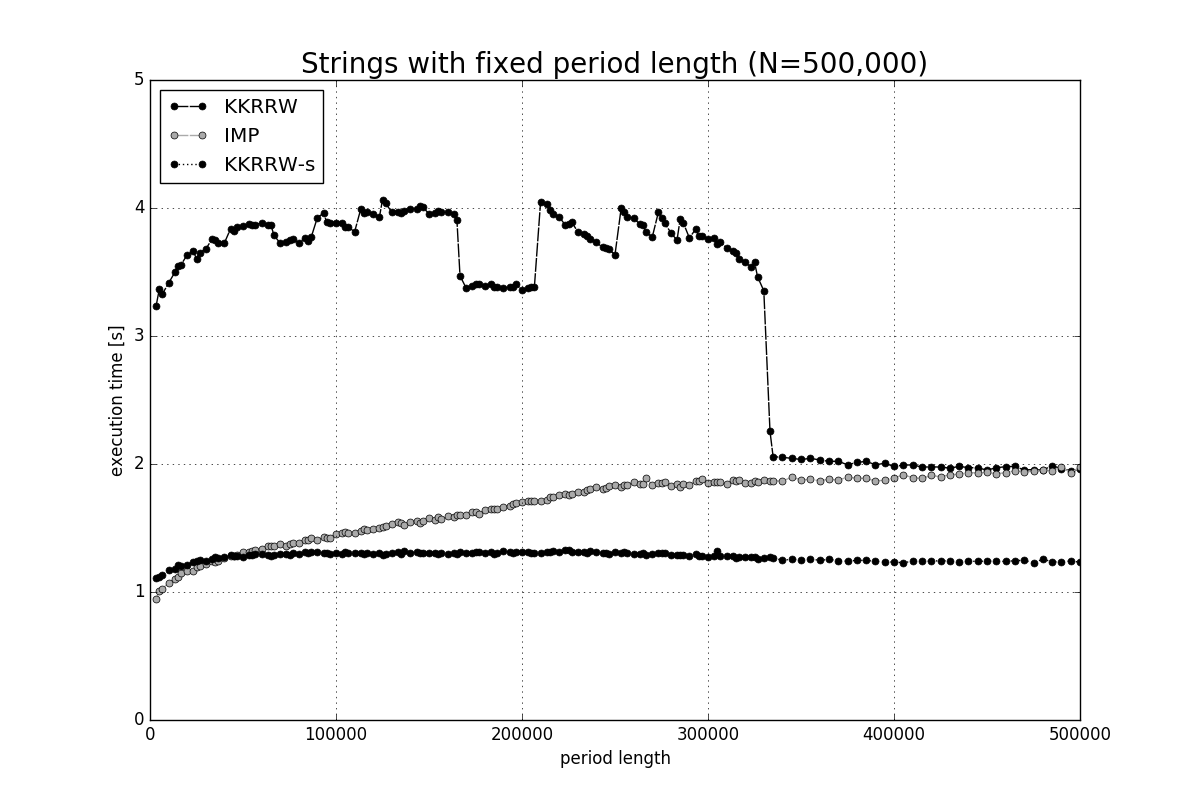}
\end{center}
\vspace*{-0.3cm}

\begin{center}
	\includegraphics[scale=0.27]{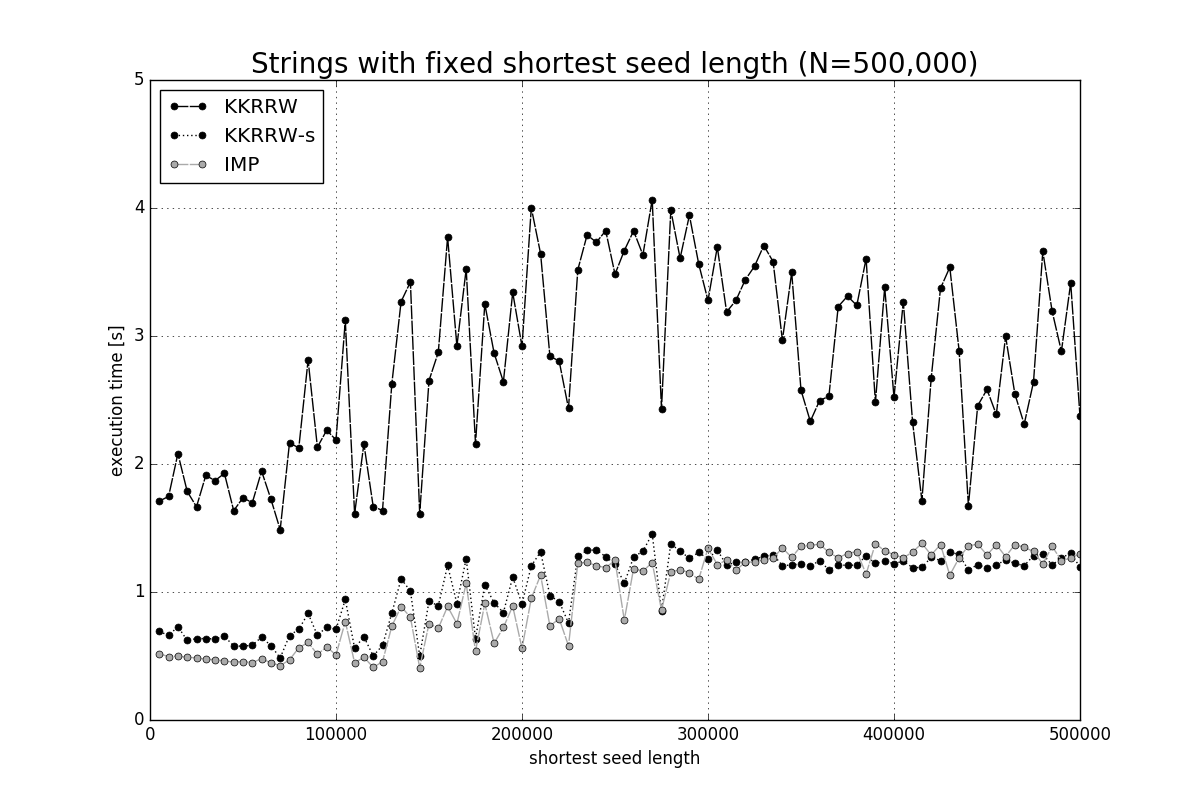}
\end{center}

\section{Joinable Segment Trees}\label{sec:JST}
A \emph{segment tree} (see, e.g., \cite{DBLP:conf/spire/RubinchikS17}), also called a static range tree,
is a data structure used for maintaining operations on sequences of a fixed length $N$
in $\Oh(\log N)$ time. As a toy example, it allows to change a given element of the sequence and report the sum
of a fragment of a sequence.

To create a segment tree, we choose the smallest power of 2, denoted as $B$ (the base), which is greater than or equal to $N$.
Then we build a full binary tree with $B$ leaves. Consecutive leaves represent consecutive elements
of the sequence. Last $B-N$ leaves remain unused.
A node of the tree which is not a leaf, $v$, accumulates information about a sequence fragment represented by leaves
in its subtree. We will call it \emph{the fragment of $v$}. In the example that was mentioned above,
$v$ stores the sum of elements of its fragment.
The values stored in nodes should be easy to update after changing the value in a leaf.
In the toy example, we could replace the value in every node on
the path from the changed leaf to the root (in this order) using the sum of values stored in the children of this node.

The figure below shows an example segment tree for $N=6$.

\begin{center}
\begin{tikzpicture}[level distance=1.5cm,
    level 1/.style={sibling distance=6cm},
    level 2/.style={sibling distance=3cm},
    level 3/.style={sibling distance=1.5cm}]

    \node {1+2+3+4+5+6}
      child {
        node {1+2+3+4}
        child {
          node {1+2}
          child { node{1} }
          child { node{2} }
        }
        child {
          node {3+4}
          child { node{3} }
          child { node{4} }
        }
      }
      child {
        node {5+6}
        child {
          node {5+6}
          child { node{5} }
          child { node{6} }
        }
        child {
          node {unused}
          child { node{unused} }
          child { node{unused} }
        }
      }
      ;
\end{tikzpicture}
\end{center}

  \subsection{Definition}
By a \emph{joinable segment tree} (JST) we mean a segment tree with memory optimization, that is, we will not
store nodes which were not initialized with any value, and add them dynamically when they are necessary. Below
there is an example of a JST with $N=6$ in which only the values in the leaves 1, 2 and 6 were initialized.
\begin{center}
\begin{tikzpicture}[level distance=1cm,
  level 1/.style={sibling distance=2cm},
  level 2/.style={sibling distance=1.5cm}]

  \node {1+2+6}
    child {
      node {1+2}
      child {
        node {1+2}
        child { node{1} }
        child { node{2} }
      }
    }
    child {
      node {6}
      child {
        node {6}
        child { node{6} }
      }
    }
    ;
\end{tikzpicture}
\end{center}

If we have two JSTs, $T_A$ and $T_B$, containing disjoint sets of initialized leaves, we can join them into
a JST $T$ containing leaves from both $T_A$ and $T_B$ using the method shown below:

{\small
\begin{verbatim}
def join(v_A, v_B): // nodes of T_A and T_B
    if v_A is None:
        return v_B
    if v_B is None:
        return v_A

    l' = join(left children of v_A and v_B)
    r' = join(right children of v_A and v_B)

    assign (l', r') as left and right children of v_A
    update(v_A)
    return v_A
\end{verbatim}
}

\noindent
The function \texttt{update} recomputes the value in $v_A$ based on values in its children.

For example, joining a JST with leaves 1, 2, 6 with a JST with leaves 3, 5:
\begin{center}
\begin{tikzpicture}[level distance=1cm,
  level 1/.style={sibling distance=2cm},
  level 2/.style={sibling distance=1.5cm}]

  \node {1+2+6}
    child {
      node {1+2}
      child {
        node {1+2}
        child { node{1} }
        child { node{2} }
      }
    }
    child {
      node {6}
      child {
        node {6}
        child { node{6} }
      }
    }
    ;
\end{tikzpicture}
\qquad
\begin{tikzpicture}[level distance=1cm,
  level 1/.style={sibling distance=1.5cm}]

  \node {3+5}
    child {
      node {3}
      child {
        node {3}
        child { node{3} }
      }
    }
    child {
      node {5}
      child {
        node {5}
        child { node{5} }
      }
    }
    ;
\end{tikzpicture}
\end{center}

\noindent
gives the following result (updated nodes are \underline{underlined}):

\begin{center}
\begin{tikzpicture}[level distance=1cm,
  level 1/.style={sibling distance=3cm},
  level 2/.style={sibling distance=1.5cm}]

  \node {\underline{1+2+3+5+6}}
    child {
      node {\underline{1+2+3}}
      child {
        node {1+2}
        child { node{1} }
        child { node{2} }
      }
      child {
        node{3}
        child {node{3}}
      }
    }
    child {
      node {\underline{5+6}}
      child {
        node {\underline{5+6}}
        child { node{5} }
        child { node{6} }
      }
    }
    ;
\end{tikzpicture}
\end{center}

\subsection{Complexity}
The total time complexity of creating and joining $N$ JSTs, $i$-th of which containing only the $i$-th leaf
initialized, is $\Oh(N\log N)$ and does not depend on the order of joins.

Indeed, let us denote the length of the fragment of node $v$ by $F(v)$. In a full binary tree,
the subtree of $v$ would contain exactly $2F(v)-2$ edges.
If during a \texttt{join} operation node $v$ was updated, it means that at least one edge was added in its subtree.
This means that during all \texttt{join}
operations node $v$ will be updated at most $2F(v)-2$ times.
In total all nodes with equal $F(v)$ value will be updated
$\frac{B}{F(v)}(2F(v)-2)=\Oh(B)=\Oh(N)$ times. There are $\log B$ different values of $F(v)$,
thus the total time complexity of all \texttt{join} operations is $\Oh(N\log N)$.

\subsection{Maxgap problem}\label{JST:maxgap}
In this section we will describe how to compute $\maxgap(\Occ_T(v))$ for every node
$v$ of the suffix tree of $T$.

For every node $v$ of the suffix tree (from the leaves to the root) we compute
a JST representing $\Occ_T(v)$. Leaf $i$ in such JST is initialized if and only if $i\in \Occ_T(v)$.
In every node $v$ of the JST we store the minimum and the maximum element of $\Occ_T(s)$ contained
in the subtree of $v$ and $\maxgap$ of all the elements from the subtree of $v$.
The function \texttt{update} will look as following:

{\small
\begin{verbatim}
def update(v):
    if one of children of v is None:
        copy values from the existing child
    else:
        v.minval = v.left.minval
        v.maxval = v.right.maxval
        v.maxgap = max(v.left.maxgap, v.right.maxgap,
                       v.right.minval - v.left.maxval)
\end{verbatim}
}

To compute $\maxgap(\Occ_T(v))$ we need to join all JSTs of the children of $v$.

\subsection{Extended Disjoint Set Union}\label{JST:FU}

Here we describe how to extend the Union operation in Disjoint Set Union
with computing the \textit{change list}, that is needed for computation of partial covers.
The Find operation is performed by a classical Find and Union structure.
Extension for Union operation is performed on Joinable Segment Trees.

At the beginning we create $n$ sets with labels $1,2,\ldots,n$.
The $i$-th set is a JST with the $i$-th leaf initialized with value $i$.
A node of JST stores the minimum and the maximum value in its subtree.
While joining trees $A$ and $B$ we will mark values from tree $A$ with color $c_A$,
and values from tree $B$ with color $c_B$.
Then, in the \texttt{update} operation we compare the maximum value of the left child ($l_{max}$) with
the minimum value of the right child ($r_{min}$). If the colors of $l_{max}$ and $r_{min}$ are different, we
add the $(l_{max}, r_{min})$ pair to the resulting change list.

As we need colors only for the \texttt{update} operation, we do not need to mark values
that are never used by it. So, if some node of the tree $A$ is relinked as a child of the tree $A\cup B$,
and we do not call \texttt{update} on it, both of its values are marked with $c_A$.
An analogous rule holds for marking values in nodes of the tree $B$ with $c_B$.
Hence, \texttt{update} copies colors along with the values.

All changes of the successor will be added to the change list, because
every two consecutive elements $i$, $j$ from $A\cup B$ will be compared in an \texttt{update} operation
on the lowest common ancestor of the leaves $i$ and $j$.

\end{document}